\documentclass[pre,preprint,superscriptaddress,showpacs,amsfonts]{revtex4}
 
\usepackage{graphicx}
\usepackage{dcolumn}
\usepackage{bm}

\usepackage{amsmath}
\begin{document}
 \title{Professor C. N. Yang and Statistical Mechanics}
\author{F. Y. Wu}
\affiliation{Department of Physics, Northeastern University, Boston, Massachusetts 02115, U.S.A.}

\begin{abstract}
Professor Chen Ning Yang has made seminal and influential contributions in many different areas in theoretical physics. This talk focuses on his contributions in statistical mechanics, a field in which Professor Yang has held a continual interest for over sixty years. His Master's thesis was on a theory of binary alloys with multi-site interactions, some 30 years before others studied the problem. Likewise, his other works opened the door and led to subsequent developments in many areas of modern day statistical mechanics and mathematical physics. He made seminal contributions in a wide array of topics, ranging from the fundamental theory of phase transitions, the Ising model, Heisenberg spin chains, lattice models, and the Yang-Baxter equation, to the emergence of Yangian in quantum groups. These topics and their ramifications will be discussed in this talk.
\end{abstract}


\maketitle
 

\section{Introduction}\label{Introduction:sec1}
Statistical mechanics is the subfield of physics that deals with systems consisting
of large numbers of particles.  It provides a framework for relating macroscopic
properties of a system, such as the occurrence of phase transitions, to microscopic
properties of individual atoms and molecules.

The theory of statistical mechanics was founded by Gibbs (1834-1903)
who based his considerations
on earlier works of Boltzmann (1844-1906) and Maxwell (1831-1879).
 By the end of the 19th century, classical mechanics was fully developed and applied successfully
 to rigid body motions.  However, after it was recognized that  ordinary materials consist of
$10^{23}$ molecules, it soon became apparent
that the application of traditional classical mechanics is fruitless
in explaining physical phenomena on the basis of  molecular considerations.
To overcome this difficulty, Gibbs proposed a statistical theory
for computing bulk properties of real materials.

Statistical mechanics as proposed by Gibbs applies to all physical systems
regardless of their macroscopic states.  But
in early years there had   been doubts about whether it could
fully explain
physical phenomena  such as phase transitions.
In 1937, Mayer \cite{mayer37} developed the method  of cluster expansions
for analyzing the statistical mechanics
of a many-particle system which worked  well for
systems in the gas phase.  This offered some hope
of explaining phase transitions, and the Mayer theory
subsequently became
the main frontier of statistical mechanical research.
   This was  unfortunate in hindsight since, as Yang and Lee would
later show (see  Sec. \ref{phasetransition:sec4}),
the grand partition function   used in the Mayer theory
cannot be continued into the condensed phase, and hence it does not settle
   the question it set out to answer.

This was the stage and status of
statistical mechanics in the  late 1930's when Professor C. N. Yang
entered college.

\section{A Quasi-chemical Mean-field Model of Phase Transition}
In 1938, Yang entered the National Southwest Associate University, a university
formed jointly by  National Tsing Hua University, National Peking University
and Nankai University during the Japanese invasion, in Kunming, China.
As an undergraduate student Yang attended seminars given by J. S. (Zhuxi) Wang,
who had recently returned from Cambridge, England, where he had
studied theory of phase transitions under R.  H. Fowler. These lectures brought C. N. Yang
in contact with the Mayer theory and other latest developments
in statistical mechanics \cite{yang90,yang88,yang2002}.

After obtaining his B.S. degree in 1942,  Yang continued to work on
an M.S. degree  in
1942-1944, and he chose
to work in statistical mechanics
under the direction of J. S. Wang. His Master's thesis included
 a study of phase transitions
using a quasi-chemical method of analysis, and  led to the publication
 of his first paper \cite{1945}.

In this paper, Yang generalized  the quasi-chemical theory of Fowler and
 Guggenheim \cite{FG} of  phase transitions in a binary alloy
to encompass 4-site interactions.   The idea of introducing multi-site
interactions to a statistical mechanical model was novel and new.
 In contrast, the first mentioning of a  lattice model
 with multi-site interactions was by
  myself \cite{wu1972} and by Kadanoff and Wegner \cite{KW1972} in 1972 -
that the 8-vertex model solved by Baxter \cite{baxter71}
is also  an Ising model with 4-site interactions.
Thus, Yang's quasi-chemical analysis of a binary alloy,
  an Ising model in disguise, predated the important
 study of a similar nature by Baxter in modern-day statistical mechanics by  three decades\,!

\section{Spontaneous Magnetization of the Ising Model}
The two-dimensional Ising model was solved by Onsager in 1944 \cite{onsager44}.
In a legendary footnote of a conference discussion, Onsager \cite{Onsager}
announced without proof a formula of the spontaneous magnetization of
the two-dimensional Ising model with nearest-neighbor interactions $K$,
\begin{equation}
I = \big( 1 - \sinh ^{-4} 2K \big)^{1/8}. \label{magnetization}
\end{equation}

Onsager never published his derivation since, as related by him later,
he had made use of some unproven results on Toeplitz determinants which he did not feel
comfortable to put in print. Since the subject matter was close to his Master's thesis,
 Yang had  studied the Onsager paper  extensively and attempted to derive (\ref{magnetization}).
But the Onsager paper was full of twists and turns offering very few clues
  to the computation of the spontaneous magnetization \cite{yang90page12}.

A simplified
version  of the  Onsager solution by Kauffman \cite{kauffman49} appeared
in 1949.  With the new insight to Onsager's solution, Yang immediately
realized that the spontaneous magnetization $I$ can be computed as an
off-diagonal matrix element of Onsager's transfer matrix.
This started Yang on the most difficult and the longest calculation
 of his career \cite{yang90page12}.

 After almost 6 months of hard work off and on,
Yang eventually succeeded in deriving the expression (\ref{magnetization})
and published the details in 1952 \cite{yang1952}.
Several  times during the course of the work, the
calculation stalled and Yang almost gave up, only to have it picked
 up again days later with the discovery of
new tricks or twists \cite{yang90page12}.
It was a most formidable {\it tour de force}  algebraic calculation
in the history of statistical mechanics.

\subsection{Universality of the critical exponent $\beta$}
At Yang's suggestion,
C. H. Chang \cite{chang1952} extended Yang's analysis of the spontaneous
magnetization to the Ising
model with anisotropic interactions $K_1$ and $K_2$, obtaining the expression
\begin{equation}
I = \big( 1 - \sinh ^{-2} 2K_1\sinh ^{-2} 2K_2 \big)^{1/8}. \label{magnetization1}
\end{equation}
This expression exhibits
the same critical exponent $\beta = 1/8$ as in the isotropic case, and marked
the first ever recognition of universality of
critical exponents, a fundamental principle of critical phenomena proposed
  by Griffiths 20 years later \cite{griffithsuniv}. 

\subsection{An integral equation}
A key step in Yang's evaluation of the spontaneous
magnetization is the solution of an integral equation
(Eq. (84) in Ref. [13])
 whose kernel is a product of 4 factors I, II, III, and IV.
Yang pioneered the use of Fredholm integral equations in the theory
of exactly solved models (see also Sec. \ref{subsec:heisenberg} below).
This particular kernel and similar ones have been used later by others, as
 they also occurred   in various forms in studies of
 the susceptibility \cite{barou} and the
$n$-spin correlation function of the Ising model
 \cite{randommatrix,ab1,ab2}.

\section{Fundamental Theory of Phase Transitions}\label{phasetransition:sec4}
As described in the above,
the frontier of statistical mechanics in the 1930's
focused on   the Mayer theory and the question whether the theory
 was applicable to all phases of a matter.
 Being thoroughly versed in the Mayer theory as well as
 the Ising lattice gas, Yang
     investigated this question  in collaboration with T. D. Lee.
   Their investigation resulted in two fundamental papers on the theory
of phase transitions \cite{yanglee52, leeyang52}.

In the first paper \cite{yanglee52}, Yang and Lee examined the question whether
 the cluster expansion  in the Mayer theory
can apply to  both the gas and condensed
phases.
This led them to  examine the convergence
of the grand partition function series in the thermodynamic limit, a question
that had not previously been closely investigated.
To see whether a single equation of state can describe different phases,
they looked at  zeroes
of the grand partition function in the complex fugacity plane,
again a consideration
that   revolutionized the study
of phase transitions.
Since an analytic function is defined by its zeroes,
 under this picture the onset of phase transitions
is signified by the pinching of zeroes on the real axis.
  This shows  that  the Mayer cluster expansion, while working well in the gas phase,
 cannot be analytically continued, and hence does not apply, in the condensed phase.
 It also rules out any possibility in describing
 different phases of a matter
     by a single equation of state.

In the second paper \cite{leeyang52}, Lee and Yang applied the principles formulated
in the first paper to the example of an Ising lattice gas.
By using the spontaneous magnetization result (\ref{magnetization}),
  they  deduced the exact
 two-phase region of the liquid-gas transition.
  This established without question that the Gibbs statistical
mechanics holds in all phases of a matter.
The analysis also led to the discovery of the remarkable Yang-Lee circle theorem, which states
 that
zeroes of the grand partition function of a ferromagnetic Ising lattice gas
always lie on a unit circle.

These two papers have  profoundly influenced modern-day
statistical mechanics as described in the following:

\subsection{The existence of the thermodynamic limit}
 Real physical systems typically consists of $N \sim 10^{23} $ particles confined
in a volume $V$. In applying Gibbs statistical mechanics to
real systems  one  takes the thermodynamic (bulk) limit $N,V\to \infty$
with $N/V$ held constant, and
 implicitly assumes that
such a limit exists. But in  their study of phase transitions \cite{yanglee52},
 Yang and Lee  demonstrated the necessity of a closer examination
 of this assumption.
 This  insight initiated  a host of rigorous studies
 of a similar nature.

The first comprehensive  study was by Fisher \cite{fisher64} who,
on the basis of earlier works of van Hove \cite{vanhove} and Groeneveld \cite{groeneveld},
established in 1964 the
existence of the bulk free energy  for systems with short-range interactions.
 For  Coulomb systems with long-range interactions the situation is more subtle, and
 Lebowitz and Lieb   established the bulk limit by making  use
of the Gauss law unique to Coulomb systems \cite{LL69}.
The existence of the bulk free energy for  dipole-dipole interactions
was subsequently established by Griffiths \cite{griffithsbulk}.
These rigorous studies led to a series of later studies on
 the fundamental question  of the stability of matter  \cite{lieb76}.

\subsection{The Yang-Lee circle theorem and beyond}
The consideration of Yang-Lee zeroes of the Ising model opened a new
window in statistical
mechanics and mathematical physics.
  The study of Yang-Lee zero loci  has
been extended to Ising models of arbitrary spins
\cite{griffiths}, to vertex models \cite{suzukifisher},
and to numerous other spin systems.

While the Yang-Lee circle theorem concerns  zeroes of the grand partition function,
 in 1964 Fisher \cite{fisherzero}  proposed to consider
zeroes of the partition function,  and demonstrated that
they also  lie on circles.
The Fisher argument  has since been made rigorous  with
the density of zeroes explicitly computed by Lu and myself \cite{luwu98,luwu01}.
The partition function zero consideration  has also been extended to the Potts model
by numerous authors \cite{chenhuwu}.

The concept of considering zeroes has also proven to be useful in mathematical physics.
A well-known intractable problem in combinatorics is the problem
of solid partitions of an integer \cite{macmahon}.
But a study of the zeroes of its generating function by Huang and myself \cite{huangwu} shows
they tend toward a unit circle as the integer becomes larger.
 Zeroes of the Jones polynomial in knot theory have also been computed, and
found to  tend toward the unit circle as the number
of nodes increases \cite{wuwang}. These findings appear to pointing to some
unifying truth lurking
beneath  the surface of many unsolved problems in mathematics and mathematical physics.

\section{The Quantization of Magnetic Flux}
During a visit to Stanford University in 1961, Yang was asked by W. M. Fairbank
whether or not the quantization of magnetic flux, if found, would be a new physical 
principle.
The question arose at a time when Fairbank and B. S. Deaver were in the middle of an experiment 
investigating the possibility of magnetic
flux quantization in superconducting rings. Yang, in collaboration with N. Byers,
 began to ponder over the question \cite{quan,quan1}.

By the time Deaver and Fairbank  \cite{fairbank} 
successfully concluded from their experiment 
that the magnetic flux is indeed quantized, Byers and Yang \cite{yangbyers} have
also reached the conclusion that the quantization result did not indicate a new property.
Rather, it can be deduced from usual quantum statistical mechanics. This was the
``first true understanding of flux quantization" \cite{quantization}.

\section{The Off-Diagonal Long-Range Order}
The physical  phenomena  of superfluidity and superconductivity
have been among the least-understood macroscopic quantum phenomena occurring in nature.
The practical and standard explanation has been based on
 bosonic considerations:  the Bose condensation in superfluidity and
 Cooper pairs in the BCS theory of superconductivity. But there had been
no understanding of a fundamental nature in substance. That was the question Yang pondered
 in the early 1960's \cite{yang90page54}.

In 1962, Yang published a paper
  \cite{odlro}  with the title  {\it Concept of
off-diagonal long-range order and the quantum phases of liquid helium and
of superconductors}, which crystallized
his thoughts on  the essence of
superfluidity and superconductivity. While
 the long-range order in the condensed phase in a real system
 can be understood, and computed,  as the diagonal element of the two-particle
density matrix,   Yang proposed in this paper
 that the quantum phases of superfluidity and
superconductivity are manifestations of
 a long-range order  in off-diagonal elements of the
density matrix. Again, this line of thinking and interpretation was totally new,
 and the paper has remained to be one that Yang
 has ``always been  fond of\," \cite{yang90page54}.

\section{The Heisenberg Spin Chain and the 6-vertex Model}
After the publication of the paper on the long-range off-diagonal order,
Yang  experimented using the Bethe ansatz  in constructing a Hamiltonian
 which can actually produce the
off-diagonal long-range order \cite{yang90page63}.  Instead,
this effort led to ground-breaking works on the Heisenberg spin chain,
the 6-vertex model,
and the one-dimensional delta function gas described below.

\subsection{The Heisenberg spin chain}\label{subsec:heisenberg}
 In a series of definitive papers in
collaboration with C. P. Yang \cite{yangyang,yangyang3}, Yang studied  the one-dimensional
Heisenberg spin chain with the Hamiltonian
\begin{equation}
H = -\frac 1 2 \sum\big( \sigma_x\sigma_x' +\sigma_y\sigma_y'
+ \Delta \sigma_z\sigma_z'\big)\, .  \label{heisenberg}
\end{equation}
 Special cases of the Hamiltonian had been considered before by others.
But Yang and Yang analyzed the Bethe ansatz
solution of the eigenvalue equation of (\ref{heisenberg})
 with complete mathematical rigor, including the rigorous
analysis of a Fredholm integral equation arising in the theory
in the full range of
$\Delta$. The  ground state energy is found to be singular
at $\Delta = \pm 1$.  Furthermore, this series of papers has become
important as it formed the
basis of ensuing studies of the 6-vertex model, the
one-dimensional delta function gas and numerous other related problems.

\subsection{The 6-vertex model}
In 1967, Lieb \cite{lieb} solved the residual entropy problem of
square ice, a prototype of the two-dimensional
6-vertex model, using the method of Bethe ansatz. Subsequently,
the solution was  extended to  6-vertex models
in the absence of an external field \cite{lieb1,lieb2}.
These solutions share the characteristics that they
are all based on Bethe ansatz analyses involving real momentum $k$.

In the same year 1967, Yang, Sutherland and C. P. Yang \cite{sutherland}
published a solution of the general 6-vertex model in the presence
of  external fields, in which they used the Bethe ansatz with complex momentum $k$.
But  the  Sutherland-Yang-Yang paper
did not provide details of the solution.
 This led others to fill in the gap in ensuing years, often with analyses starting from
scratch,  to understand the thermodynamics.
 Thus, the $\Delta < 1$ case was studied by Nolden \cite{nolden}, the
$\Delta \geq 1$ case by Shore and Bukman \cite{shore,shore1}, and the case $|\Delta| = \infty$
by myself in collaboration with
Huang {\it et al.} \cite{huangetal}  The case of $|\Delta| = \infty$ is of particular interest,
since it is also a 5-vertex model as well as  an honeycomb lattice dimer model
with a nonzero dimer-dimer interaction.  It is
the only known soluble  interacting close-packed dimer model.

\section{One-dimensional Delta Function Gases}
\subsection{The Bose gas}
The first successful application of the Bethe ansatz to a many-body
problem was  the one-dimensional delta function Bose
gas solved by Lieb and Liniger \cite{liebdelta,liebdelta1}. Subsequently, by extending considerations
to include all excitations,
Yang and C. P. Yang deduced
the thermodynamics of the  Bose gas
\cite{yangyang1}.  Their theoretical prediction has
 recently been found
to agree very well with experiments on a one-dimensional
Bose gas trapped on an atom chip \cite{yangyang2}.

\subsection{The Fermi gas}
The study of the delta function Fermi gas was
more subtle.
In a seminal work having profound and influential impacts
in many-body theory, statistical mechanics and mathematical physics,
Yang in 1967 produced the full solution of the delta function Fermi gas \cite{yang67}.
The solution was obtained as a result of the combined use of group theory
 and the nested Bethe ansatz, a repeated use of
the Bethe ansatz devised by Yang.

One very important ramification of the Fermi gas work is the exact solution
of the ground state of the one-dimensional Hubbard model
obtained by Lieb and myself \cite{Liebwu,Liebwu1,Liebwu2}.   The solution of the Hubbard model
is similar to that of the delta function gas except with the replacement
of the momentum $k$ by $\sin k$ in the Bethe ansatz solution.
Due to its relevance in high $T_c$ superconductivity, the Lieb-Wu solution
has since led to a torrent of further works on the one-dimensional Hubbard model
 \cite{korepin}.

\section{The Yang-Baxter Equation}
The two most important integrable models in statistical mechanics
are the delta function Fermi gas solved by Yang \cite{yang67} and the 8-vertex model
solved by Baxter \cite{baxter71, baxter80}.
The key to the solubility of
 the delta function gas
 is an operator relation \cite{yang671} of the S-matrix,
\begin{equation}
{Y_{jk}}^{ab}\,{Y_{ik}}^{bc}\,{Y_{ij}}^{ab}= {Y_{ij}}^{bc}\,{Y_{ik}}^{ab}\,{Y_{jk}}^{bc},
   \label{YBE1}
\end{equation}
and for the 8-vertex model the key is a relation \cite{baxter801}
of the 8-vertex operator,
\begin{equation}
U_{i+1}(u)U_{i}(u+v)U_{i+1}(v)=U_{i}(v)U_{i+1}(u+v)U_{i}(u).
\label{YBE2}
\end{equation}
Noting the similarity of the two relations and realizing they are fundamentally
the same, in a paper on the 8-vertex model Takhtadzhan and Faddeev \cite{tf}
called it the Baxter-Yang relation.
Similar relations also arise in other quantum and lattice models.
These relations have since been  referred to as
the {\it Yang-Baxter equation} \cite{perk, perk1}.

The Yang-Baxter equation is
an internal consistency condition among parameters in a quantum or lattice model,
and can usually be written down
by considering a star-triangle relation \cite{perk,perk1}.
  The solution of the Yang-Baxter equation, if found,
often aids in solving the model itself.
 The Yang-Baxter equation has been shown to play a
 central role  in connecting many  subfields
in mathematics, statistical mechanics and mathematical physics \cite{ge}.

\subsection{Knot invariants}
One example of the role played by the Yang-Baxter equation in mathematics is
 the construction of
knot (link) invariants. Knot invariants are algebraic
quantities, often in polynomial forms, which preserve topological
properties of three-dimensional knots.
In the absence of definite prescriptions, very few
knot invariants were known for decades. The situation changed dramatically
after the discovery of the Jones polynomial  by Jones in 1985 \cite{jones}.
  and the subsequent
revelation that knot invariants can be constructed from lattice models in
statistical mechanics \cite{kauffman}.

The key to constructing knot invariant from statistical mechanics is the
Yang-Baxter equation. Essentially, from each lattice model whose Yang-Baxter
equation possesses a solution, one  constructs a knot invariant.
 One example is  the Jones polynomial which can be constructed from a solution
of the Yang-Baxter equation of the Potts model, even though the solution is
in an unphysical regime \cite{kauffman88}.   Other examples are described in a 1992
 review  on knot theory and statistical mechanics by myself \cite{wu92}.

\subsection{The Yangian}
In 1985, Drinfeld \cite{drinfeld} showed that there exists a Hopf
algebra (quantum group)
over $SL(n)$ associated with the Yang-Baxter equation (\ref{YBE1})
after the operator $Y$ is expanded into a series.  Since Yang found the
  first rational solution of the expanded equation,  he named the Hopf
algebra the {\it Yangian} in honor of Yang \cite{drinfeld}.

Hamiltonians
with the Yangian symmetry include, among others, the one-dimensional
Hubbard model, the delta function Fermi gas, the Haldane-Shastry model \cite{haldane},
and the Lipatov model \cite{lipatov}.
The Yangian algebra is of increasing importance in
quantum groups, and has been used very recently in a formulation of  quantum
entangled states \cite{GeMolin}.

\section{Conclusion}
In this talk I have   summarized the contributions made by Professor Chen Ning Yang
in  statistical mechanics. It goes without saying that
it is not possible to cover all aspects of Professor Yang's
work in this field, and undoubtedly there are omissions.
But it is clear from what is presented, however limited, that Professor C. N. Yang
has made immense contributions to this relatively young
field of theoretical physics.

A well-known  treatise in statistical mechanics is the
20-volume  {\it Phase Transitions and
Critical Phenomena} published in   1972 - 2002 \cite{series,series1}.
The series covers almost every subject matter of traditional statistical mechanics.
The first chapter of Volume 1 is an introductory  note by Professor Yang,
in which he assessed the  status of the field
and  remarked about possible future directions of statistical mechanics.

At the conclusion he wrote:

{\it One of the great intellectual challenges for the next few decades}

{\it is the question of brain organization.}

 \noindent
As research in biophysics and brain memory functioning  has mushroomed
into a major field in recent years,
 this is an extraordinary prophecy and a testament to
  the insight and foresight of Professor Chen Ning Yang.

\section{Acknowledgments}
I would like to thank Dr. K. K. Phua for inviting me to the Symposium.
I am grateful to M.-L. Ge  and J. H. H. Perk for  inputs in the preparation of
the talk, and to
J. H. H. Perk for a critical reading of the manuscript.

\end{document}